\documentclass[onecolumn]{article}
 \usepackage{authblk}
 
\usepackage[right]{lineno}
\usepackage{amsmath,amsfonts}
\usepackage{url}
\usepackage{acronym}
\usepackage{siunitx}
\usepackage{subcaption}
\usepackage{eurosym}
\usepackage{multirow}
\usepackage{graphicx}
\usepackage{a4wide}

\usepackage{array}
\newlength\savedwidth

\newcommand\thickhline{\noalign{\global\savedwidth\arrayrulewidth\global\arrayrulewidth 2pt}%
	\hline
	\noalign{\global\arrayrulewidth\savedwidth}}

\newacro{ANOVA}[ANOVA]{analysis of variance}
\newacro{ECA}[ECA]{embodied conversational agent}
\newacro{HMD}[HMD]{head-mounted display}
\newacro{HRTF}[HRTF]{head-related transfer function}
\newacro{IHTA}[IHTA]{Institute for Hearing Technology and Acoustics}
\newacro{SEM}[SEM]{Standard Error of Mean}
\newacro{STM}[STM]{short-term memory}
\newacro{SUS}[SUS]{Slater, Usoh, and Steed presence questionnaire}
\newacro{VA}[VA]{Virtual Acoustics}
\newacro{VR}[VR]{virtual reality}
\newacro{VRLab}[VRLab]{virtual reality laboratory}
\newacro{VSR}[VSR]{verbal serial recall}
\newacro{avVSR}[avVSR]{audiovisual verbal serial recall}
\newacro{DV}[DV]{dependent variable}
\newacro{IV}[IV]{independent variable}
\newacro{IV_SP}[IV\_SP]{serial position}
\newacro{IV_Angle}[IV\_Angle]{audiovisual angle incongruence}
\newacro{IV_Disp}[IV\_Disp]{display device}
\newacro{IV_Voice}[IV\_Voice]{audiovisual voice incongruence}
\newacro{CI}[CI]{Credible Interval}
\newacro{HDI}[HDI]{Highest Density Interval}
\newacro{PPD}[PPD]{Posterior Probability Distribution}
\newacro{PD}[PD]{probability of direction}
\newacro{ROPE}[ROPE]{region of practical equivalence}
\newacro{SD}[SD]{standard deviation}
\newacro{elpd}[elpd]{expected log pointwise predictive density}
\newacro{LOO}[LOO]{Leave-one-out}

\begin{document}

	\begin{flushleft}
		{\Large
			\textbf\newline{Audiovisual angle and voice incongruence do not affect audiovisual verbal short-term memory in virtual reality} 
		}
		\newline
        Audiovisual verbal short-term memory under angle and voice incongruence in virtual reality
        \newline
		\\
		Cosima A. Ermert\textsuperscript{1,*},
        Manuj Yadav\textsuperscript{1},
        Jonathan Ehret\textsuperscript{2},
		Chinthusa Mohanathasan\textsuperscript{3},
		Andrea Bönsch\textsuperscript{2},
		Torsten W. Kuhlen\textsuperscript{2},
		Sabine J. Schlittmeier\textsuperscript{3},
		Janina Fels\textsuperscript{1}
		\\
		\bigskip
		\textbf{1} Institute for Hearing Technology and Acoustics, RWTH Aachen University, Aachen, Germany
		\\
		\textbf{2} Visual Computing Institute, RWTH Aachen University, Aachen, Germany
		\\
		\textbf{3} Work and Engineering Psychology, RWTH Aachen University, Aachen, Germany
		\\
		\bigskip

		* cosima.ermert@akustik.rwth-aachen.de (CE)
		
	\end{flushleft}

\section*{Abstract}
\Ac*{VR} environments are frequently used in auditory and cognitive research to imitate real-life scenarios, presumably enhancing state-of-the-art approaches with traditional computer screens. 
However, the effects of different display technologies on audiovisual processing remain underexplored.
This study investigated how \acs*{VR} displayed with an \ac*{HMD} affects serial recall performance compared to traditional computer monitors, focusing on their effects on audiovisual processing in cognitive tasks.
For that matter, we conducted two experiments with both an \acs*{HMD} and a computer monitor as display devices and two types of audiovisual incongruences: angle (Exp. 1) and voice (Exp. 2) incongruence.
To quantify cognitive performance an \ac*{avVSR} task was developed where an \ac*{ECA} was animated to speak the target digit sequence.
Even though subjective evaluations showed a higher sense of presence in the \acs*{HMD} condition, we found no effect of the display device on the proportion of correctly recalled digits.
For the extreme conditions of angle incongruence in the computer monitor presentation the proportion of correctly recalled digits increased marginally, presumably due to raised attention, but the effect is likely too small to be meaningful.
Response times were not affected by incongruences in either display device across both experiments.
These findings suggest that the \acs*{avVSR} task is robust against angular and voice audiovisual incongruences, irrespective of the display device, at least for the conditions studied here.
Hence, the study introduces the \acs*{avVSR} task in \acs*{VR} and contributes to the understanding of audiovisual integration.

\section*{Introduction}
\label{sec:introduction}
Experiments in auditory and cognitive sciences research are increasingly conducted in \ac{VR} environments~\cite{parsons.VirtualRealityEnhanced.2015, pieri.VirtualRealityTechnology.2023}.
\ac{VR} can evoke a higher feeling of presence, the feeling of "being in a scene"~\cite{wilkinson.MiniReviewPresence.2021}, compared to computer monitor presentations~\cite{gorini.RoleImmersionNarrative.2011}.
Thus, \ac{VR} is widely employed to create immersive environments that approximate real-life scenarios.
However, the effect of \ac{VR} on participants is not yet fully understood and appears to be highly dependent on the experimental design.

Several studies have reported positive effects of \ac{VR}, such as increased attention~\cite{li.EnhancedAttentionUsing.2020, breuer.ExaminingAuditorySelective.2022} or enhanced motivation or engagement~\cite{wenk.EffectImmersiveVisualization.2023, huang.MotivationEngagementPerformance.2021}. 
However, studies have also reported disadvantages arising from using \ac{VR}, since it can cause cybersickness~\cite{weech.PresenceCybersicknessVirtual.2019} and evoke a higher mental load~\cite{magosso.EEGAlphaPower.2019,redlinger.EnhancedCognitiveTraining.2021}.

Compared to traditional lab setups with computer monitors, in most cases, experiments in \ac{VR} differ in terms of the visual information presented.
It has been shown that auditory and visual information can be integrated into a single multimodal percept~\cite{vatakis.CrossmodalBindingEvaluating.2007}.
However, incongruences between auditory and visual stimuli have been shown to hinder multimodal integration, both with and without \ac{VR}.
For example, spatially separated and hence incongruent auditory and visual stimuli can be integrated as a single percept if the separation angle is small enough~\cite{chen.IntersensoryBindingSpace.2013,thurlow.CertainDeterminantsVentriloquism.1973}.
This \emph{ventriloquism effect} has been replicated in \ac{VR}~\cite{huisman.VentriloquistEffectNot.2022}.
A large angular separation, however, can impede user experience in terms of presence, immersion, and cybersickness, while small separation angles are tolerated in \ac{VR}~\cite{kim.StudyingEffectsCongruence.2022}.
Semantic audiovisual incongruence, i.e., seeing an object (such as a "salmon") and hearing an incongruent description (such as "pasta") results in stronger brain responses compared to congruent stimuli~\cite{tromp.CombinedUseVirtual.2018}.
Incongruence in lip movement (\emph{McGurk effect}~\cite{mcgurk.HearingLipsSeeing.1976}) can be detected more easily in \ac{VR} than in computer monitor settings~\cite{siddig.PerceptionDeceptionAudioVisual.2019}.
Providing coherent user interaction in \ac{VR}, e.g., by synchronized audiovisual virtual hand illusions~\cite{choi.MultisensoryIntegrationVirtual.2016} or sound feedback according to the user's movements, e.g., walking~\cite{hoppe.VRsneakyIncreasingPresence.2019}, can increase immersion.

These studies primarily focused on perception-based effects of audiovisual incongruences, such as detectability, perceived presence, or immersion.
However, as \ac{VR} is increasingly employed in cognitive research, it is crucial to understand whether incongruences impact cognitive functions. 
While it is logical to hypothesize that noticeable audiovisual incongruences could impede cognitive task performance, to the best of our knowledge, this has not been demonstrated yet. 
Our study aims to bridge this gap by investigating the extent to which audiovisual incongruences affect short-term memory, and whether such a potential effect differs between \ac{HMD} and traditional computer monitor presentations. 
Our findings are likely to deepen the understanding of multisensory integration and audiovisual interaction while providing insights into scene design for audiovisual and cognitive research in \ac{VR}.

Two studies were carried out with both an \ac{HMD} and a computer monitor to enable a direct comparison of performance differences between these display devices.
In the first experiment, an audiovisual angle incongruence was introduced, while participants were presented with a voice incongruence in the second experiment.
To measure short-term memory, a serial recall task~\cite{hurlstone.MemorySerialOrder.2014} is employed. 
In \ac{VSR} tasks participants have to remember a sequence of verbal items for a short period of time.
The verbal items can be presented visually (as images~\cite{delogu.SemanticEncodingWorking.2009}, written~\cite{surprenant.IrrelevantSpeechPhonological.1999, harvey.InputOutputModality.2007}, or as lip-movements~\cite{divin.EffectsIrrelevantSpeech.2001}), or auditorily (as spoken words~\cite{schlittmeier.DoesIrrelevantMusic.2008, surprenant.IrrelevantSpeechPhonological.1999}, like in a listening situation).
For a discussion of the differences between visual and auditory digit presentations, please refer to Yadav et al.~\cite{yadav.CognitivePerformanceOpenplan.2023}.
For \ac{VR} applications and to be able to integrate incoherences, this unimodal task had to be extended to audiovisual presentation in the present paper. 
In previous studies, when audiovisual presentation has been chosen for memory tasks, it has usually involved showing an image of the item and playing back a corresponding sound or word at the same time~\cite{delogu.SemanticEncodingWorking.2009, vollmer.StimulusOnsetAsynchronies.2022}.
This differs from real-life audiovisual stimuli: when a person is speaking, the auditory information is accompanied by spatial information and visual cues like lip movement, gaze, and gestures~\cite{theze.AnimatedVirtualCharacters.2020a}.
Therefore, in this study, we created an audiovisual version of the serial recall task in \ac{VR}, where specific attention was paid to providing realistic auditory and visual stimuli.
For that matter, the audio stimulus was the spoken digit sequence which contained spatial information and the visual stimulus was an animated \ac{ECA} speaking the target sequence.
This task will be called \ac{avVSR} in the following.
We also used questionnaires to link performance data in terms of the proportion of correctly recalled digits and response times to subjective evaluations, such as the perceived presence.

\section*{Experiment 1: Audiovisual angle incongruence}
In selecting the audiovisual incongruence for Experiment 1, our objective was to manipulate an attribute inherent to the target signal, so that the participants would naturally focus on it, while simultaneously ensuring that their ability to successfully complete the task was not impeded.
The latter could occur if the participant is presented with two competing digit streams, such as hearing a "three" and seeing a digital image or lip movement of a five.
In the coherent case the information is presented bimodally, whereas in the incoherent case, only unimodal information can be used while the other modality has to be ignored actively (e.g.,~\cite{hughes.ImpactOrderIncongruence.2005}).
To avoid such imbalances in the task design, we sought to introduce incongruences in a task-irrelevant attribute of the target signal. 

Given that in realistic scenes, auditory and/or visual sources almost always possess a directional component, i.e., they are perceived at a certain distance and angle from the listener, we decided to introduce an incongruence in the angle from which the auditory and visual sources are perceived to create a discrepancy between the auditory and visual scenes.
In this experiment, digit sequences were played back using spatial audio reproduction and an \ac{ECA} was animated to speak the sequence simultaneously.
An audiovisual angle incongruence ranging from $0\si{\degree}$ to $60\si{\degree}$ was introduced between the position of the auditory source and the position of the visual source, the \ac{ECA}. 

The task was performed with both \ac{HMD} and computer monitor presentations and answers and response times were logged.
Both display devices were evaluated with questionnaires with respect to the perceived presence. 

\subsection*{Method}
\subsubsection*{Participants}
A total of 25 adults were recruited for the experiment via mailing lists and posters in the authors' institutes between 27th of April and 13th of May 2022. Participants were required to be native German speakers (as the entire experiment was conducted in German), have normal hearing, and (corrected-to-)normal vision. 
Normal hearing below $25\,\si{\dB}$ SPL~\cite{who.ReportInformalWorking.1991} between $250\,\si{Hz}$ and $14\,\si{\kilo\hertz}$ was tested with an AURITEC Ear 3.0 audiometer and Sennheiser HDA300 headphones using a pulsed pure tone ascending audiometry. 
Two participants failed the audiometry and had to be excluded. 
$N=23$ participants (16 male, 6 female, 1 non-binary, aged between 19 and 36 years, $M = 25.74$, $SD=4.03$) passed the audiometry. 
Normal or corrected-to-normal vision was validated with a Snellen chart up to (20/30)~\cite{snellen.ProbebuchstabenZurBestimmung.1873}. 
All participants gave written informed consent before the experiment began. 
Participants received a $10\,$\euro{} gift voucher for a bookstore. 
The experiment procedure was pre-approved by the Ethics Committee at the Medical Faculty of RWTH Aachen University (EK396-19). 
The study was conducted in accordance to the rules of conduct stated in the Declaration of Helsinki.

\subsubsection*{Apparatus and materials}
\begin{figure}[h!]
\centering
    \begin{subfigure}[t]{.45\textwidth}
        \centering
       \includegraphics[width=.9\linewidth]{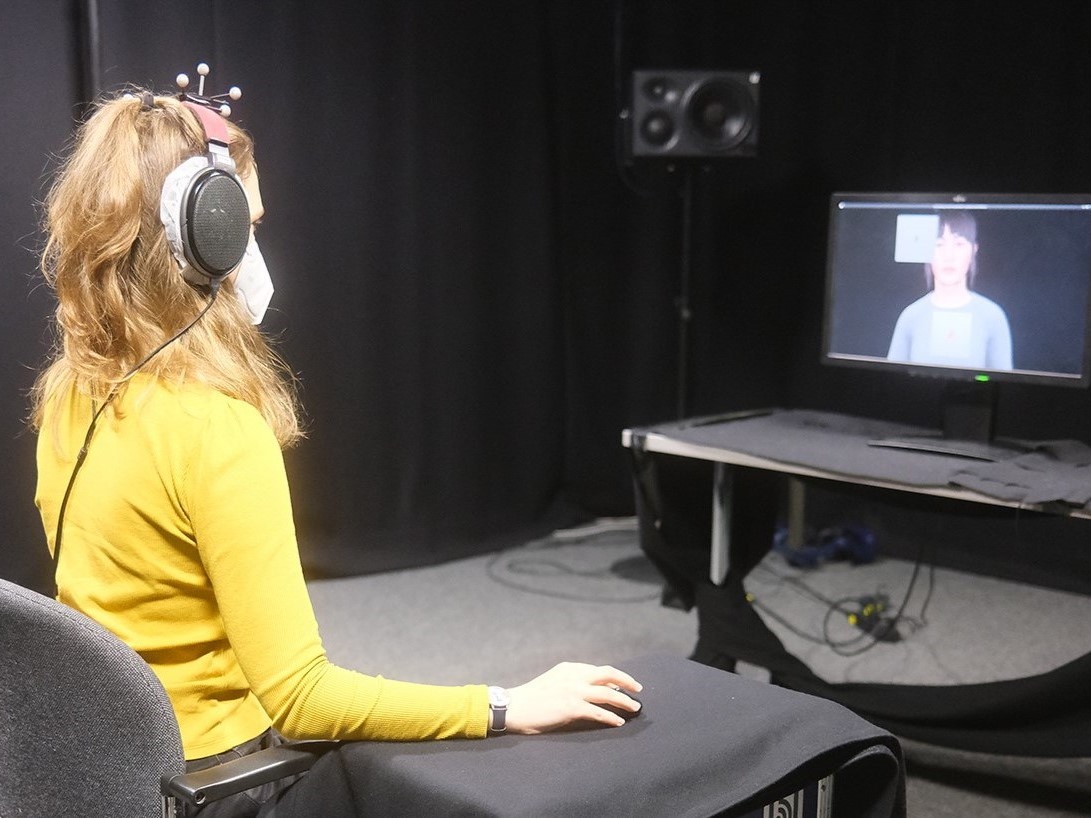}
        \caption{Computer monitor presentation. Participant is wearing headphones with Optitrack tracking markers.}
        \label{fig:computermonitorsetup}
    \end{subfigure}%
    \vspace{0.3cm}
    \begin{subfigure}[t]{.45\textwidth}
        \centering
        \includegraphics[width=.9\linewidth]{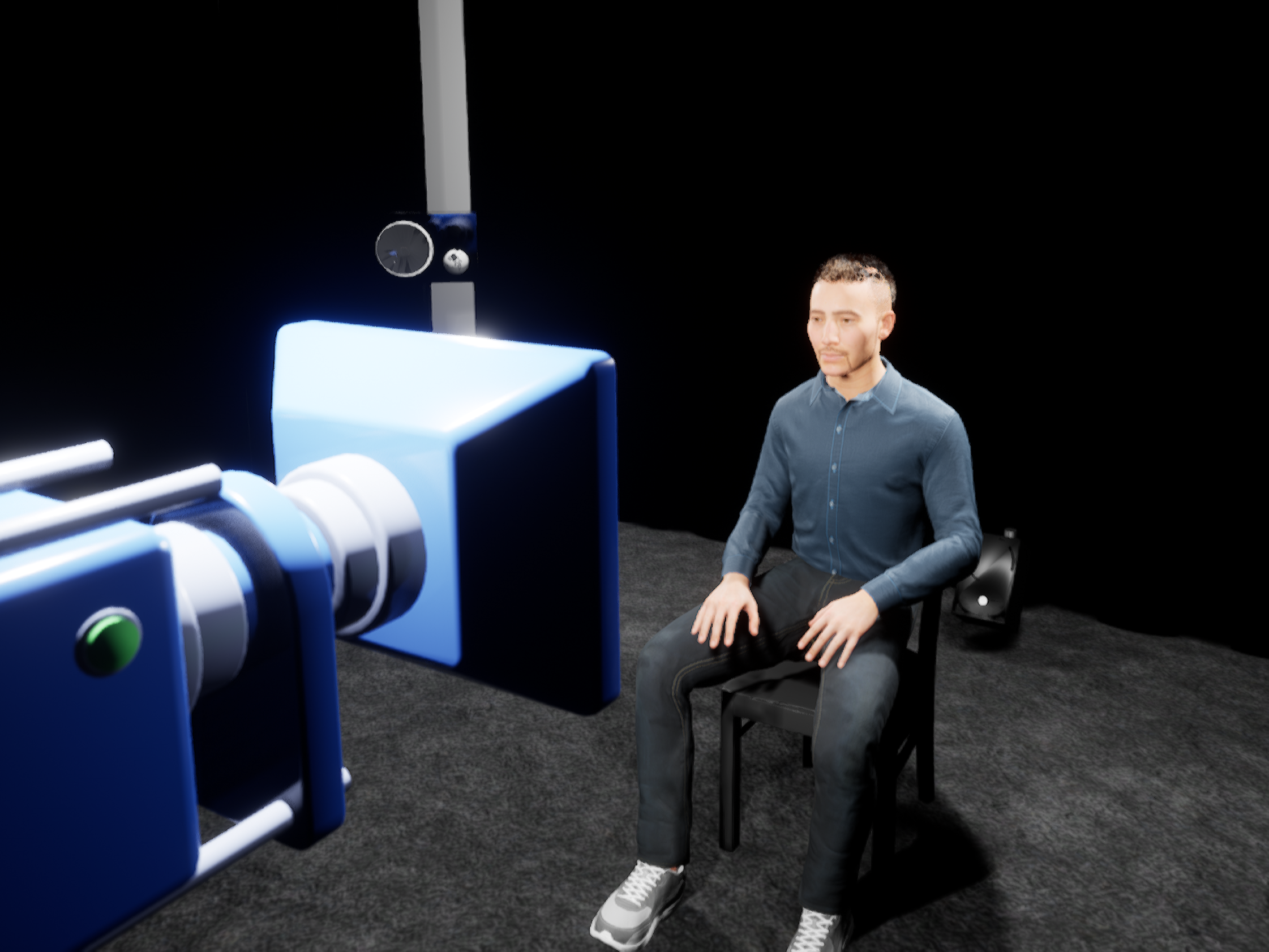}
        \caption{\ac{VR} scene created in Unreal Engine. The blue camera represents the participants' view.}
        \label{fig:vrsetup}
    \end{subfigure}
    \caption{Listening experiment setup. The female \ac{ECA} is employed in Experiment 1 and 2. The male \ac{ECA} is employed only in Experiment 2.}
    \label{fig:ExperimentSetup}
\end{figure}

The \ac{avVSR} task was configured for both \ac{HMD} and computer monitor presentations and implemented in Unreal Engine (Epic Games, v4.27) using the following plugins: the StudyFramework plugin~\cite{ehret.StudyFrameworkComfortablySetting.2024} (v4.26), which handles the data logging and experiment procedure, the RWTH \ac{VR} Toolkit plugin~\cite{gilbert_2024_10817754} (v4.27) for basic \ac{VR} interaction, and the Character plugin (v4.27) for \ac{ECA} animation.
In the \ac{HMD} condition, the scene was presented using an HTC Vive Pro Eye (dual AMOLED screens with 3.5$''$ diagonal display, resolution $1440\times1600$ pixels per eye) and one HTC Vive controller. In the computer monitor condition, a Fujitsu B24T-7 (24$''$ diagonal display, resolution $1920\times1080$ pixels) monitor and a wireless computer mouse (Cherry MW2310) were used. 
To ensure consistency between the presentation modalities, the \ac{VRLab}~\cite{pausch.DocumentationExperimentalEnvironments.2022} of the \ac{IHTA}, in which the experiment took place, was replicated in \ac{VR}, so that the participants were ``in the same room'' regardless of the display device used (see Fig.~\ref{fig:ExperimentSetup}). 

For the visual stimulus, a female \ac{ECA} (see Fig.~\ref{fig:computermonitorsetup}) was created with MetaHuman Creator (Epic Games, v0.5.2).
Using Oculus Lip Sync (Meta, v20),  lip animation was generated from the audio files such that the \ac{ECA} moved its lips according to the digit sequence heard. 
In the \ac{HMD} condition, the \ac{ECA} sat on a virtual chair at a distance of $d = 2.5\,\si{\meter}$ from the participant, which is a comfortable inter-person distance to an unknown person~\cite{hall.HiddenDimensionMan.1969, bonsch.SocialVRHow.2018}. 
In the computer monitor condition, only the upper part of the body was visible on the computer monitor. 
This image section and the distance of the computer monitor from the participants were chosen so that the size and position of the \ac{ECA} were consistent between the two display devices.  
Participants sat on a chair, which was adjusted so that the participants (measured from the top of the head) and the \ac{ECA} were approximately at the same height. 
The two main visual differences between display devices were that in the \ac{HMD} condition the participants could not see themselves and the \ac{ECA}'s lower body was not visible in the computer monitor condition.

The acoustic digit stimuli were presented in German and taken from the database by Oberem and Fels~\cite{oberem.SpeechMaterialParadigm.2020} (voice \textit{female b}). 
The original publication did not include a recording of the digit five, but was made available by the authors.
No loudspeakers were used to avoid visual fixation to possible source positions. 
Instead, virtual sound sources for a binaural headphone-based reproduction were auralized with \ac{VA}~\cite{institute_for_hearing_technology_and_aco_2024_13744474} (v2020a) in a reflection-free (anechoic) environment using dynamic scene rendering. 
Communication between Unreal Engine and \ac{VA} was established via the Virtual Acoustics Plugin (v4.26). 
This was achieved by tracking the head movements of the participants and virtually shifting the sound source position accordingly. 
Head-tracking was done in two different ways depending on the display device: for the \ac{HMD}, the built-in tracking system was used.
This was not possible in the computer monitor presentation. 
Instead, an OptiTrack system with Motive (OptiTrack, v2.1.1) tracked head movements. 
To provide plausible localization cues, the \ac{HRTF} of the \ac{IHTA} artificial head~\cite{schmitz.NeuesDigitalesKunstkopfmesssystem.1995} was employed. 
As shown by Oberem et al.~\cite{oberem.ExperimentsLocalizationAccuracy.2020}, this generic \ac{HRTF} provides sufficiently accurate localization in dynamic scenes for most listeners.

The auditory stimuli were played back using Sennheiser HD650 open-back headphones with a Behringer ADA8200 sound card for both display devices. 
All auditory stimuli were calibrated to $60\,\si{\dB}$(A) as the power sum of both ears using the \ac{IHTA} artificial head~\cite{schmitz.NeuesDigitalesKunstkopfmesssystem.1995}.
Headphone equalization was performed individually for each participant following Masiero and Fels~\cite{masiero.PerceptuallyRobustHeadphone.2011}.

\subsubsection*{Audiovisual incongruence}
\begin{figure*}[ht!]
	\centering
	\includegraphics[width=.55\textwidth]{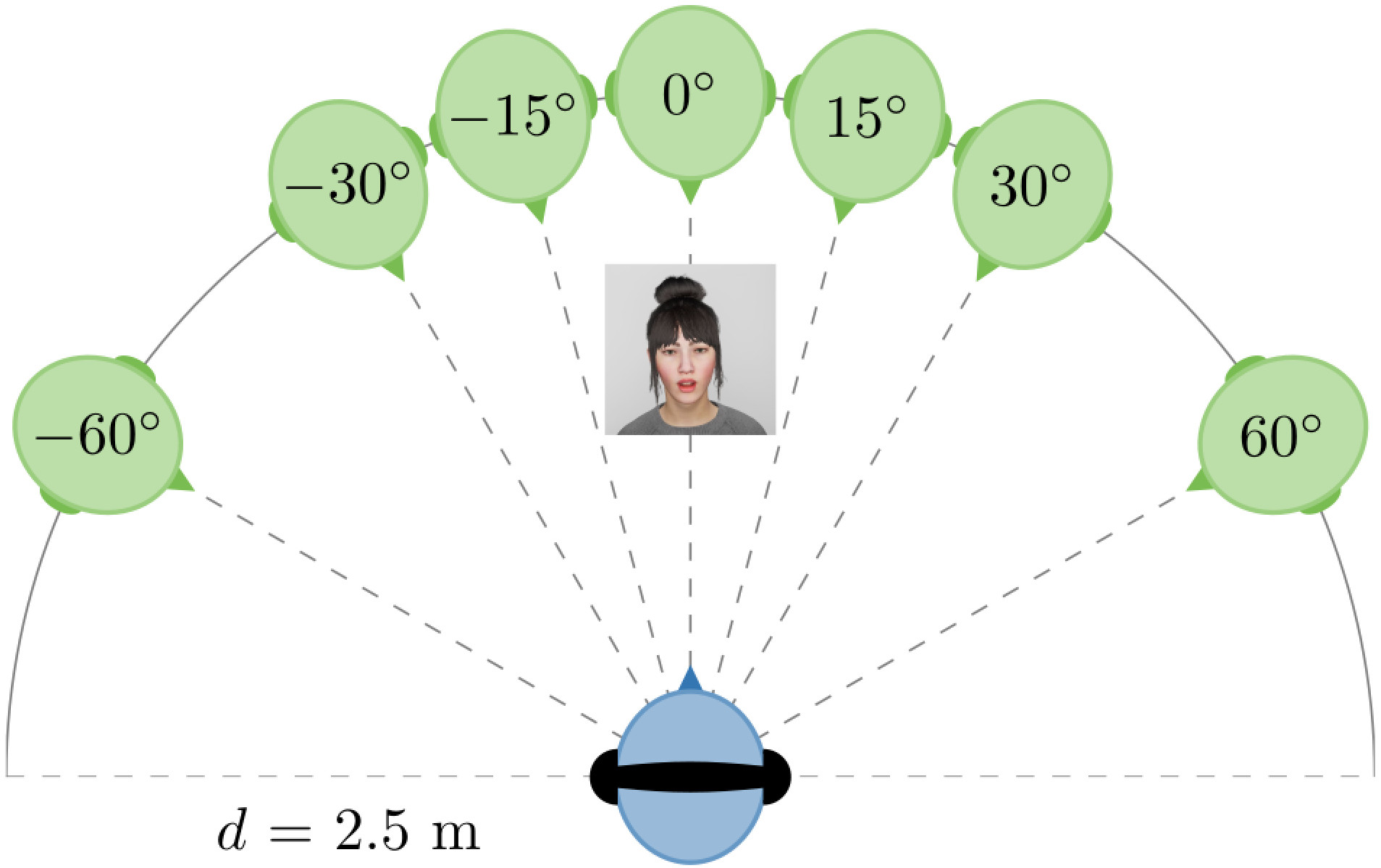}
	\caption{{\bf Audiovisual angle incongruence in Experiment 1.} The female \ac{ECA} is displayed at an azimuth angle of $0^\circ$ on the horizontal plane at a distance of $d = 2.5\,\si{\metre}$ from the listener (blue). The possible virtual sound source positions (green) are at the azimuth angles $0^\circ,\pm15^\circ,\pm30^\circ,$ and $\pm60^\circ$ at the distance $d$.}
	\label{fig:Exp1_ang}
\end{figure*}
The position of the \ac{ECA} as the visual source was kept constant at an azimuth angle of $\varphi = 0^\circ$, straight ahead of the participant (see Fig.~\ref{fig:Exp1_ang}). 
Audiovisual angle incongruence should include angles small enough to allow the audiovisual stimulus to be perceived as a unit and angles large enough so that an audiovisual angle incongruence is expected to be detectable.
Since the conversational partners are approximately at the same height in seated settings, we limited our selection to angles on the horizontal plane.
Studies on the ventriloquism effect report different threshold angles up to which spatially separated visual and auditory stimuli are integrated. 
It is suggested that audiovisual angle incongruence should not exceed $15^\circ$ for audiovisual integration to occur~\cite{slutsky.TemporalSpatialDependency.2001, chen.IntersensoryBindingSpace.2013}, but integration effects up to $30^\circ$ have been shown on the horizontal plane~\cite{jack.EffectsDegreeVisual.1973}. 
In their study on audiovisual congruence in \ac{VR}, Kim and Lee~\cite{kim.StudyingEffectsCongruence.2022} found that participants generally tolerate audiovisual angle mismatches up to $30^\circ$ on the horizontal plane and detect angle offsets of $15^\circ$ only with a probability of $18.8\%$.
It should be noted that these angular limits are usually determined with simplified audiovisual stimuli such as noise bursts and flashing lights. 
Thresholds for realistic sounds, such as speech, have hardly been investigated~\cite{huisman.VentriloquistEffectNot.2022}.

Since headphone reproduction with a generic \ac{HRTF} is used, a localization offset compared to a headphone reproduction with an individual \ac{HRTF} would be expected.
However, Oberem et al.~\cite{oberem.ExperimentsLocalizationAccuracy.2020} showed that this localization offset is non-significant on the horizontal plane with dynamic reproduction with head movements for the given generic \ac{HRTF} with speech stimuli.

Based on these findings, four audiovisual angle incongruences were chosen. 
As a baseline condition, $\Delta\varphi = 0^\circ$ was included, where the position of the auditory target and visual target are aligned. 
Furthermore, $\Delta\varphi = \pm 15^\circ$, $\pm 30^\circ$, and $ \pm 60^\circ$ were included with decreasing probability of integration expected for increasing angular differences.

\subsubsection*{Questionnaires}
\label{sec:quest}
To link the objective performance in the display devices with subjective impressions and to quantify presence, two distinct questionnaires were used: one questionnaire that had to be completed for each display device and a second general questionnaire. 
Both questionnaires were implemented using SoSciSurvey in German~\cite{leiner.SoSciSurvey.2019}.

In the first questionnaire, the perceived presence during the \ac{HMD} presentation compared to the computer monitor presentation was evaluated with a German translation~\cite{boensch.SocialWayfindingStrategies.2024} of the \ac{SUS}~\cite{usoh.UsingPresenceQuestionnaires.2000} on a 7-point Likert scale with the corresponding anchors.
Additionally, participants were asked about the appearance of the \ac{ECA} (following Ehret at al.~\cite{ehret.ProsodyEmbodimentInfluence.2021a}): (\textbf{Q\_Nat}: \textit{How natural did the interlocutor seem to you?}, \textbf{Q\_Spe}: \textit{How natural did the way the interlocutor spoke seem to you?}, \textbf{Q\_Sen}: \textit{To which degree did the interlocutor appeared to be a sentient being?}). Furthermore, the modality participants focused on (\textbf{Q\_Vis}: \textit{How much did you focus visually on the interlocutor?}) and the controller handling (\textbf{Q\_Con}: \textit{How intuitive was the handling of the controller?}) were examined.

At the end of the experiment, participants completed a second questionnaire. 
This questionnaire focused on the task itself (\textbf{Q\_Dif}: \textit{How difficult was the task for you?}) and on the audiovisual incongruence (\textbf{Q\_Inc}: \textit{Did you notice any audiovisual incongruence?} [Asked on a yes/no basis], \textbf{Q\_Imp}: \textit{If you noticed the audiovisual incongruence, did it impact your performance?}, \textbf{Q\_Dom}: \textit{If you noticed the audiovisual incongruence, did you shift your attention towards one domain (audio, visual) as a result?}). 
All these ratings, except for \textbf{Q\_Inc}, were made on a 7-point Likert scale between 1 (minimum, e.g., "Not at all") and 7 (maximum, e.g., "Strongly").
A comment field was provided for additional remarks.

\subsubsection*{Procedure}
The experiment was conducted in individual sessions at the \ac{IHTA}, RWTH Aachen University.
The audiometric and visual screening was performed in a sound isolated hearing booth~\cite{pausch.DocumentationExperimentalEnvironments.2022}.
After passing the screening, the main experiment started in the \ac{VRLab}~\cite{pausch.DocumentationExperimentalEnvironments.2022}. 
The display device was varied across participants in two blocks, meaning that participants had to complete all trials on one display device before moving to the next one. 
The starting display device was counterbalanced between participants.
At the beginning of the experiment, written instructions explaining the task were given. 
Participants were instructed not to vocalize the digits or use their fingers as an aid. 
The instructions were followed by a training block consisting of eight trials, two per audiovisual angle incongruence in counterbalanced order. 
The training was followed by twelve trials for each of the audiovisual angle incongruence ($|\Delta\varphi| = {0^\circ, 15^\circ, 30^\circ, 60^\circ}$), yielding a total of $12\times4=48$ trials per display device in the main experiment, divided into three blocks of 16 trials each. 
Each angle incongruence was presented twice in a block, once from the left and once from the right side; the angle incongruence $|\Delta\varphi|=0^\circ$ was presented twice from the front. 
The order of angle incongruences was counterbalanced across participants to avoid order effects. Between blocks, participants could take a break of no fixed length. 
After each display device block, participants filled out the first questionnaire.

\begin{figure*}[htb!]
	\centering
	\includegraphics[width=\textwidth]{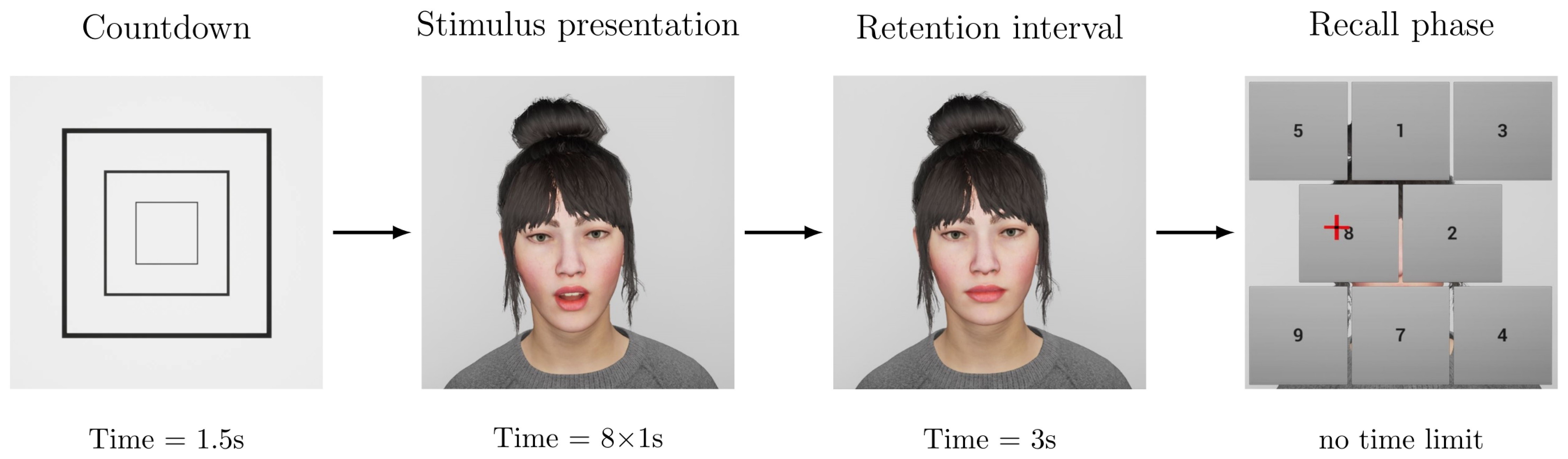}
	\caption{{ \bf One trial of the \ac{avVSR} task.} Graphic depiction of the trial phases \textit{Countdown}, \textit{Stimulus presentation}, \textit{Retention interval}, and \textit{Recall phase} over time. In the \textit{Recall phase}, the cursor is visible as a red cross.}
	\label{fig:trial}
\end{figure*}
Each experimental trial was structured as follows (see Fig.~\ref{fig:trial}): after a visual countdown consisting of three rectangles decreasing in size at a rate of one rectangle per $500\,\si{\milli\second}$, rendered on a plane which covered the \ac{ECA} entirely, participants saw the \ac{ECA} speak a sequence of eight digits at a rate of one digit per second.
The digit sequences were created from digits between one and nine in such a way that no digit was repeated and no more than two consecutive steps, up or down, occurred (e.g., "1-2-3-4" or "9-8-7-6" would not be allowed).  
After the stimulus presentation, participants had to wait for a retention interval of three seconds before they were asked to reproduce the order of the digits by clicking on the corresponding fields in a matrix, which was rendered as blocks in front of the \ac{ECA} (see Fig.~\ref{fig:trial}, Recall phase).
During this recall phase, the \ac{ECA} was partially visible and animated with an idle movement.
Only the eight digits present in the sequence were displayed in the matrix. 
The order of the digits within the matrix was changed randomly for each trial to avoid the use of visuo-spatial recall techniques~\cite{healy.ShortTermMemoryOrder.1982}. 
Once participants clicked on a digit, it disappeared. 
Corrections were not possible. 
The proportion of correctly recalled digits and the response times were logged.

Participants completed the second questionnaire at the end of the experiment. 
The reason for presenting the second questionnaire at the end rather than between the display device blocks was to avoid participants' performance being affected by inadvertently disclosing the research questions of the listening experiment.
The entire experiment, including screening and questionnaires, took approximately 90 minutes.

\subsubsection*{Statistical analysis}
All statistical analyses were conducted in the software R (version 4.2.2). The R package \emph{tidyverse}~\cite{Wickham_et_al} was used for data processing and plots. 
For the \ac{avVSR} task, there were two \acp{DV} (i.e., fixed effects): the \textit{Accuracy} and the average response times \textit{RT\_mean}. 
\textit{RT\_mean} is defined as the average time between two button clicks in the recall phase or, for the first digit, the time between the start of the recall phase and the first button click. 
\textit{Accuracy} is the proportion of digits recalled at the correct serial position. 
The \acp{IV} included the \ac{IV_SP}, the \ac{IV_Angle} (see Fig.~\ref{fig:Exp1_ang}), and the \ac{IV_Disp}. 
Separate Bayesian generalized mixed-effects models were created for each \ac{DV} using the R package \emph{brms} (version 2.18)~\cite{Buerkner_2017}. 
In each model, the repeated-measures experimental design was incorporated as independently varying intercepts (i.e., random effects) for the participants. 
Since Accuracy includes proportion data with many values close or equal to 1, a zero-one-inflated Beta distribution with a logit link function was used as the family for the models with this \ac{DV}. 
A Gamma distribution with a log link function was used as the family for \textit{RT\_mean} as the DV since such distributions are commonly used for time data.

The modeling process for the \ac{avVSR} task included starting with an intercept-only baseline model, followed by including an \ac{IV} or an interaction between the \acp{IV} in subsequent models. 
The prior distributions per \ac{IV} and their interactions were sampled 16000 times, using four independent chains of 5000 samples each and discarding the first 1000 warm-up samples. 
To avoid dependencies per chain, No-U-Turn Sampling was employed. Weakly-informative prior distributions (in the logit or log scale, depending on the \ac{DV}) were used, which for the fixed effects included normal distributions each with a mean (\emph{M}) of 0 and a \emph{\ac{SD}} of 2. The latter is relatively large given the scales, and hence conservative. For the random effects, heavily tailed Student-t distributions with \emph{M} = 0 and \emph{\ac{SD}} = 2 each were used. The performance across models was compared using the \ac{LOO} cross-validation criteria, based on the difference in values of the \ac{elpd} and the standard error of the \ac{elpd} of the models~\cite{Sivula_Magnusson_Matamoros_Vehtari_2023}. Posterior predictive checks were performed for all the models.

Pairwise comparisons across \ac{IV} levels were conducted using the R package \emph{emmeans}~\cite{emmeans}. These comparisons are summarized using the median and 95\% \acp{CI} calculated using the \ac{HDI} of the \ac{PPD}. The Bayesian 95\% \ac{CI} provides the interval within which 95\% of the \ac{PPD} lies. This interval considers the probability range or the uncertainty/belief regarding the parameter value (e.g., mean), contingent on both the present sample and the prior information. This presents a more direct statement regarding the parameter than the 95\% confidence interval used in frequentist statistics. The latter generally means that upon repeated random sampling and calculating the 95\% confidence interval per sample, the true parameter value would be contained within approximately 95\% of those intervals. This so-called long-run probability takes all values being equally likely a priori and parameter values are then contingent upon the present sample.  

To determine whether an effect (comparison across \ac{IV} levels) exists, we consider the \ac{PD} of the PPD, which provides the probability of an effect going in either the positive or negative direction. \ac{PD} further provides a link to frequentist statistics in that the two-sided \emph{p}-values of .05 and .01 approximately correspond to \ac{PD} of 97.5\% and 99.5\%, respectively. \ac{PD}, however, is not an index of significance. In order to consider whether an effect is meaningful/significant, we use the proportion of the 95\% \ac{CI} inside the \ac{ROPE}. The latter is the range signifying an effect of negligible magnitude and is conceptually similar to the null hypothesis in frequentist statistics. The range for the \ac{ROPE} was specified as $\pm0.1 \times \emph{SD}_{IV}$~\cite{kruschke2018bayesian}. \ac{PD} and \ac{ROPE} were calculated using the R package \emph{bayestestR}~\cite{bayestestR}.

For each of the questionnaire items, separate Bayesian mixed-effects models were created with the \ac{IV_Disp} as the fixed effect and independently varying intercepts (random effects) for the participants. For all questions except for Q\_Inc, the ratings (the \ac{DV}) were modelled using the cumulative family with a probit link, since the ratings were on an ordinal scale. For Q\_Inc, the rating was modelled using the Bernoulli family with a logit link, since the ratings were binary (yes/no). A weakly informative prior with a mean of 0 and \emph{\ac{SD}} of 5 was used for the \acp{DV}. Other modeling criteria were the same as used for the \ac{avVSR} data. The differences in ratings between the display devices are summarized using the \emph{\ac{SD}} and 95\% \ac{CI} (calculated using the \ac{HDI} of the \ac{PPD}) for all questions except Q\_Inc, for which the change in log-odds of the response and the associated 95\% \ac{CI} is used.

\subsection*{Results}

\subsubsection*{\Acl{avVSR}}
Based on the \ac{LOO} criteria with \emph{Accuracy} as the \ac{DV}, the baseline (intercept-only) model was the worst model, while the performance across models with the various \acs{IV} and their interactions were not substantially different. Hence, the model with all the \acp{IV} and an interaction term between the \acp{IV} angle incongruence and display was selected as the final model since it encompasses the experimental design most comprehensively. The interaction term with serial position was not considered useful for the research intend and hence not included. The pairwise comparison for the final model is presented in Table~\ref{tab:Exp1_Accuracy}. The median differences across the comparisons were generally quite low with PD less than 97.5\% except for the difference between the $0\si{\degree}$ and $60\si{\degree}$ angles with the computer monitor display device. The latter indicates that there is a higher \textit{Accuracy} in the $60\si{\degree}$ compared to the $0\si{\degree}$ condition, suggesting an effect. However, the high percentage of the \ac{PPD} within the \ac{ROPE} suggests that this effect is likely of negligible practical significance, as it falls within the range which is generally considered too small to be meaningful.

For \emph{RT\_mean} as the \ac{DV}, similar to \textit{Accuracy}, the final model included all the \acp{IV} and an interaction term between the \acp{IV} audiovisual angle incongruence and display device. However, none of the pairwise comparisons were statistically meaningful with PPDs with low PD values and high \% in ROPE, and are not considered further.  

\begin{table}[]
\centering
		\caption{\bf Summary of pairwise comparisons between the audiovisual angle incongruences and display device combinations with \textit{Accuracy} as the \ac{DV}.} 
\begin{tabular}{|l|l|l|l|l|l|l|}
			\hline
\multicolumn{2}{|l|}{\bf{Comparison}}                          & \multirow{ 2}{*}{\bf{Median}}    
& \multirow{ 2}{*}{\bf{95\% CI}}           
& \multirow{ 2}{*}{\bf{PD}}      
& \multirow{ 2}{*}{\bf{\% in ROPE}} \\
\cline{1-2}
\bf{Angle} & \bf{Device} &&&&\\

\thickhline
$0\si{\degree}-15\si{\degree}$ &    Monitor  & -0.01     & {[}-0.04, 0.01{]}  & 83.88\% & 72.15\%    \\
$0\si{\degree}-30\si{\degree}$ &    Monitor   & -0.02     & {[}-0.04, 0.01{]}  & 91.64\% & 56.24\%    \\
$0\si{\degree}-60\si{\degree}$ &    Monitor   & -0.03     & {[}-0.06, -0.01{]} & 99.69\% & 11.01\%    \\
$15\si{\degree}-30\si{\degree}$ &    Monitor   & -5.13$e^{-03}$ & {[}-0.03, 0.02{]}  & 65.78\% & 89.24\%    \\
$15\si{\degree}-60\si{\degree}$ &    Monitor   & -0.02     & {[}-0.05, 0.00{]}  & 95.71\% & 45.07\%    \\
$30\si{\degree}-60\si{\degree}$ &    Monitor   & -0.02     & {[}-0.04, 0.01{]}  & 90.73\% & 61.86\%    \\
\hline
$0\si{\degree}-15\si{\degree}$ &    \ac{HMD}   & 9.44$e^{-03}$  & {[}-0.02, 0.03{]}  & 76.80\% & 81.20\%    \\
$0\si{\degree}-30\si{\degree}$ &    \ac{HMD}   & 2.77$e^{-03}$  & {[}-0.02, 0.03{]}  & 50.84\% & 93.01\%    \\
$0\si{\degree}-60\si{\degree}$ &    \ac{HMD}     & -7.33$e^{-03}$& {[}-0.03, 0.02{]}  & 72.30\% & 86.04\%    \\
$15\si{\degree}-30\si{\degree}$ &    \ac{HMD}       & -9.14$e^{-03}$ & {[}-0.03, 0.02{]}  & 77.12\% & 82.05\%    \\
$15\si{\degree}-60\si{\degree}$ &    \ac{HMD}      & -0.02     & {[}-0.04, 0.01{]}  & 91.25\% & 61.02\%    \\
$30\si{\degree}-60\si{\degree}$ &    \ac{HMD}    & -7.67$e^{-03}$ & {[}-0.03, 0.02{]}  & 72.99\% & 86.01\%   \\
\hline
\end{tabular}
\begin{flushleft} CI = Bayesian credible interval, PD = probability of direction, ROPE = region of practical equivalence 
\end{flushleft}
\label{tab:Exp1_Accuracy}
\end{table}

\subsubsection*{Questionnaire}
\begin{figure*}[]
	\centering
	\includegraphics[width=0.8\textwidth]{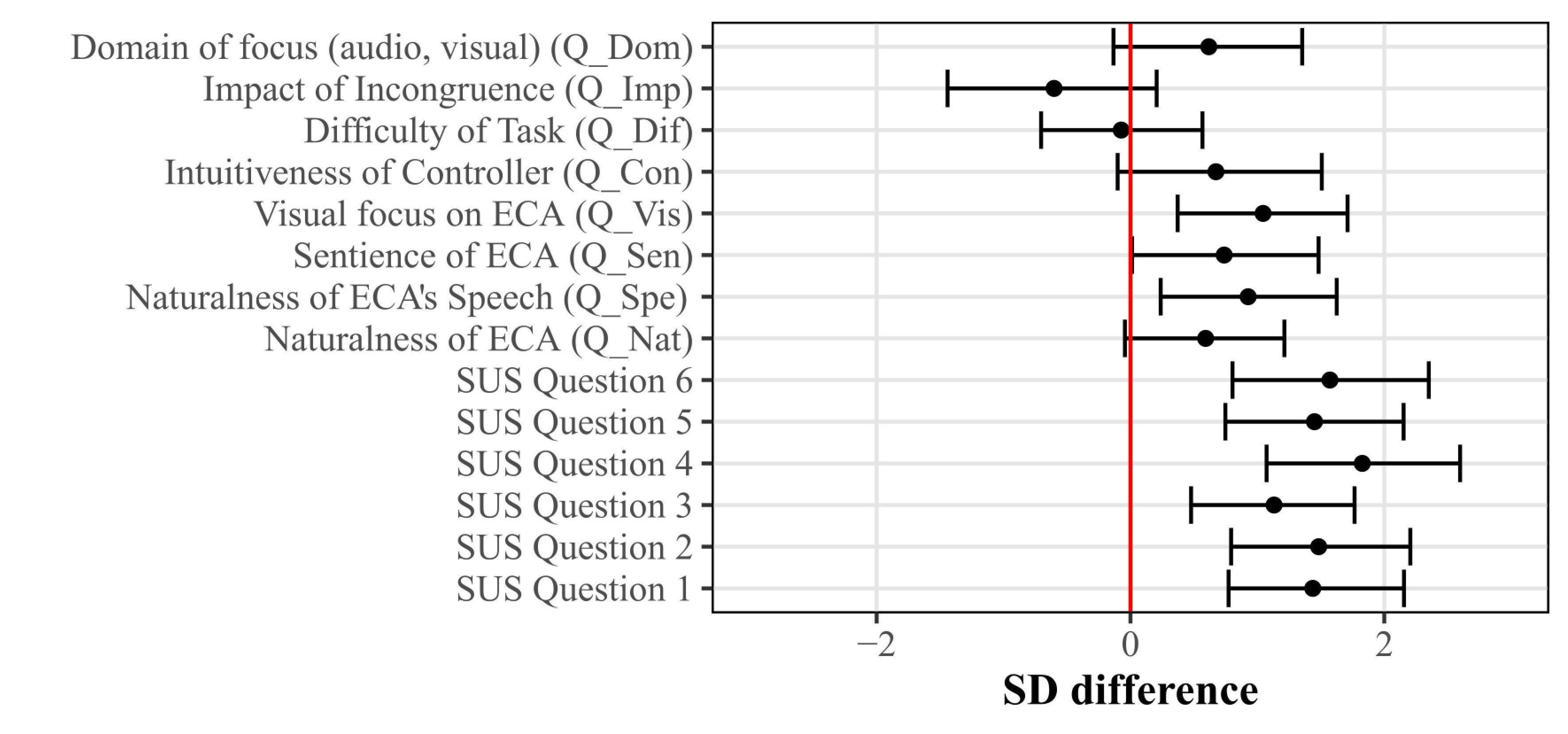}
	\caption{\textbf{Questionnaire results of Experiment 1}. The standard deviation \emph{SD} difference between the HMD and the Monitor display conditions is shown for each question on the y-axis. Error bars indicate the 95\% credible intervals. All questions were rated on a scale of 1 to 7, except for Q\_Inc, which was asked on a yes/no basis and is thus not displayed here.}
	\label{fig:Exp1_Quest}
\end{figure*}

As seen in Fig.~\ref{fig:Exp1_Quest}, for all the questions from the \ac{SUS} questionnaire, the \emph{SD} differences between the \ac{HMD} and computer monitor display conditions were statistically robust, indicating that the participants reported a higher perceived presence in the \ac{HMD} condition. This is depicted by the 95\% CI of the \emph{SD} difference between display conditions not crossing the zero line, suggesting a meaningful effect. 

For the other questions, the 95\% \ac{CI} crosses the zero line, indicating that the differences in \emph{SD} of the ratings between display devices were not statistically robust. 
This holds true for the naturalness of the \ac{ECA} (Q\_Nat), the extent to which the \ac{ECA} was perceived as a sentient being (Q\_Sen), the intuitiveness of the controller handling (Q\_Con), the perceived difficulty of the task (Q\_Dif), and the perceived impact of the incongruence on the participants' performance (Q\_Imp). 

Four of the 23 participants reported not noticing the audiovisual angle incongruence (Q\_Inc). 
The change in the log odds of the perceived audiovisual incongruence (Q\_Inc) between the monitor and HMD condition (-1.14) was not statistically robust with the 95\% CI ([-4.63, 1.74]) crossing zero.
This indicates that the perceived noticeability of the incongruence did not differ between the display devices.

Statistically robust \emph{SD} differences between ratings of the display devices could be detected for the naturalness of speech (Q\_Spe), with a higher rating of naturalness in the \ac{HMD} condition, and in the visual focus (Q\_Vis), where participants reported that they focused significantly more on the \ac{ECA} in the \ac{HMD} condition.

\subsection*{Discussion}
The participants' \ac{avVSR} performance was evaluated in terms of \textit{Accuracy} and \textit{RT\_mean} for both \ac{HMD} and computer monitor presentations. 
No statistically robust influence of the display device on either metric could be found, which is in accordance with the participants' self-assessment that the task difficulty was similar for both display devices. 
This suggests that the  \ac{avVSR} task can be successfully conducted in \ac{VR}, at least in conditions close to the given setup. 
However, while the \ac{SUS} questionnaire data indicated a significantly higher perceived presence in the \ac{HMD} condition, this increased presence did not translate into changes in cognitive task performance. 
This finding highlights that higher perceived presence in \ac{VR} does not necessarily correspond with enhanced performance in cognitive tasks.

Regarding the impact of audiovisual incongruences, a marginal increase in recall performance was observed for an angle incongruence of  $60\si{\degree}$ in the computer monitor presentation.
This is quite interesting as a contrary effect, i.e., a decrease in performance, could have been expected.
However, it has been shown before that more demanding tasks can evoke increased performance~\cite{marsh.WhyAreBackground.2018}, presumably because participants put in more effort.
Furthermore, semantically incongruent audiovisual stimuli have been shown to cause stronger brain responses (reflected in more negative amplitudes) than congruent ones~\cite{tromp.CombinedUseVirtual.2018}, indicating an unexpected audiovisual stimulus pairing.
A similar effect may have occured here, i.e., audiovisual incongruence might have had an alerting effect, leading to higher concentration and, thus, improved performance.
However, we can only speculate as to why this effect occurred in the computer monitor, but not in the \ac{HMD} condition.
As indicated in the questionnaire (Q\_Vis), participants focussed more on the visual signal, the \ac{ECA}, in the \ac{HMD} than in the computer monitor condition.
It is possible that they visual stimulus dominated over the auditory signal and thus the incongruence could be ignored better.
Furthermore, \acp{HMD} influence sound source localization~\cite{huisman.AmbisonicsSoundSource.2021} and depth perception~\cite{kroczek.InfluenceAffectiveVoice.2023}.
This could have impacted the perception of the incongruence as well.
Since the effect is rather small, further investigations are necessary to validate it.

For all other combinations of display device and angle incongruence, no effect of incongruence on recall performance and reaction time was detected.
This is consistent with participants' ratings, which indicated that they felt only marginally influenced by the audiovisual angle incongruence. 
One possible explanation for the lack of interference is that participants could largely ignore the incongruence, perhaps because the task-relevant information was primarily auditory, with visual information being redundant. 

It is worth noting that the audiovisual incongruence could also be interpreted as a visual distractor. 
There are only very few studies on the impact of visual distractors on \ac{VSR} and visual signals have not been shown to impact recall performance. 
Liebl et al.~\cite{liebl.CombinedEffectsAcoustic.2012} investigated relatively rapid changes in workspace lighting, namely the effects of a traveling beam on a projector screen in the participants' field of vision, and found no effect on the visual \ac{VSR} performance.
In an experiment with flashing color stripes during stimulus presentation, Lange~\cite{lange.DisruptionAttentionIrrelevant.2005} found no reduction in performance for visual \ac{VSR}.
To the best of the authors' knowledge, auditory \ac{VSR} has not been investigated with visual distractors yet. 
However, since the ineffectiveness of visual distractors on \ac{VSR} is argued to be due to the separation of auditory-verbal and visual-spatial short-term memory according to the model of Baddeley and Hitch~\cite{baddeley.WorkingMemory.1974}, it is reasonable to assume that the audiovisual incongruence did not act as a performance-relevant distractor in the present study with auditory reproduction either. 

In summary, our results suggest that participants are able to ignore audiovisual angle incongruences in \ac{avVSR} tasks, at least in experimental settings resembling the current.

The ratings regarding the general appearance of the \ac{ECA} did not differ between display devices in a statistically robust way.
Similar effects regarding \acp{ECA} can be found in literature.
McKeown et al.~\cite{mckeown.ComparingVirtualReality.2017} conducted a study where participants rated social interactions presented either on an \ac{HMD} or a computer monitor. 
These ratings were not influenced by the display device.
In a study on human-\ac{ECA} interactions, Zojaji et al.~\cite{zojaji.ImpactImmersivenessPersuasiveness.2023} found no influence on the perceived persuasiveness and offensiveness between display devices.
This suggests that the perception of an \ac{ECA} may not generally be influenced by the display device.
Interestingly, in the present study, \ac{ECA}'s speech was perceived as more natural in the \ac{HMD} condition compared to the computer monitor condition, although the exact same material (and in the case of the speech even audio reproduction) was used. 
This may be linked to the higher presence overall and higher visual focus reported in the \ac{HMD} condition.

The experiment had several limitations for both display devices. 
Firstly, with the employed animation technique (computer-generated lip movement, no blinking, and only idle movement), the subjective evaluation of the naturalness of the \ac{ECA} was generally low.
Based on the general comments, the participants found the \ac{ECA}'s eye movement, or rather the lack thereof, particularly unnatural, which provides the opportunity for improvement.
To better imitate real-life scenarios, the animation of the \ac{ECA} should be improved. 
Secondly, regarding the choice of incongruence, the behavioral results indicate that participants were able to ignore it.
According to the questionnaire, a total of 13\% of participants did not notice the angle incongruence at all.
It is possible that the selected angle differences were insufficiently large, and that only the $60\si{\degree}$ condition entered the range of disruptive incongruences. 
We derived the selection of angles from ventriloquism experiments, in which participants focus on the spatial characteristics of the stimuli since their task is to localize them.
In contrast, the perception of spatial direction was incidental in our study, while subjects were instructed to focus on the content, specifically the to-be-remembered digits.
This could have shifted the threshold for detection, and/or the need to attend to incongruences.
The choice of using a generic \ac{HRTF} for the reproduction might have distorted the localization.
Also, participants were likely experienced in integrating audiovisual angle incongruences.
For example, when watching a movie no directional discrepancy between audio and video is perceived actively, even though loudspeakers are usually placed beside the screen.
It remains unclear if the participants were able to ignore the incongruence due to the small angle range and the familiarity of the situation - or if incongruent audiovisual stimuli can be ignored in general if they are not task-relevant.

\section*{Experiment 2: Voice incongruence}
A limitation of Experiment 1 was that the incongruence went unnoticed by many participants. 
Moreover, the angle incongruence made it challenging to distinguish between the congruent and incongruent cases clearly, as  the angle is a continuous variable with no clear threshold for noticeability.
To address this issue, a clearly noticeable incongruence was considered in Experiment 2.
One potential approach to ensure perceptibility is a pre-test with each participant to determine individual angular thresholds of detectibility, similar to the study by Kim and Lee~\cite{kim.StudyingEffectsCongruence.2022}.
However, even with this refined approach, we could not be sure that the participants would interpret the incongruence as such.
As discussed previously, most people naturally integrate incongruent audiovisual stimuli, especially in entertainment applications.
Therefore, we wanted to use a binary and more easily distinguishable incongruence which allows for clearly comparing congruent with incongruent cases.
For this purpose, in Experiment 2, a voice swap is used as the audiovisual incongruence. 

\subsection*{Method}
\subsubsection*{Participants}
The recruitment procedure, screening, ethics approval, consent form, and compensation were the same as in Experiment 1. 
Participants were recruited between 30th of November and 23rd of December 2022.
$N = 26$ adults (16 male, 10 female) aged between 21 and 39 years ($M = 28.08,\,SD=4.39$) participated in Experiment 2.
It was not an exclusion criterion if participants had already participated in the first study. 
 
\subsubsection*{Audiovisual incongruence}
In this experiment, the position of both the auditory and visual sources was kept constant at an azimuth angle of $\varphi = 0^\circ$ on the horizontal plane. 
Two \acp{ECA}, one female and one male (see Fig.~\ref{fig:vrsetup}), were introduced to the participants with their respective voices.
In the experiment, they unexpectedly spoke with the voice of the other \ac{ECA}, thus, swapping voices for an audiovisual incongruence (male-female, female-male).
Voice swaps have been used in cognitive research before: Walker-Andres et al.~\cite{walker-andrews.InfantsBimodalPerception.1991} found that infants as young as six months prefer to look at audiovisual stimuli with congruent gender compared to incongruent gender. 
In an experiment of just noticeable onset asynchronies of audiovisual stimuli, Vataksi and Spence~\cite{vatakis.CrossmodalBindingEvaluating.2007} found a higher sensitivity for gender-incongruent stimuli. 
They suggested that a incongruence between the auditory and visual stimuli in terms of gender results in them not being perceived as a unit.
A study by Szerszen~\cite{szerszen.AudioVisualMismatch.2011} indicated that a gender incongruence in the voice leads to an uncanny valley effect. 

\subsubsection*{Apparatus and materials} 
The same equipment and software as in Experiment 1 were used. 
A second, male, MetaHuman was created.
Since participants reported that the gaze of the \ac{ECA} seemed unnatural in Experiment 1, an idle eye movement was added for both \acp{ECA} using the VHGazeComponent of the Character Plugin in Speech Mode (for more information see~\cite{ehret.WhosNextIntegrating.2023}. 
The male voice stimuli were taken from~\cite{oberem.SpeechMaterialParadigm.2020} (voice \textit{male a}).

\subsubsection*{Questionnaires}
The questionnaires remained unchanged from Experiment 1. 

\subsubsection*{Procedure}
The procedure was mainly the same as in Experiment 1.
The training block consisted of six trials (three with the female \ac{ECA}, three with the male \ac{ECA}) with matching voices.
This way, participants could familiarize themselves with the task and the voice matches.
In the main experiment, twelve trials were presented for all four possible combinations of the \acp{ECA} and their voices, resulting in $12\times4=48$ trials per display device, divided into four blocks of twelve trials each. 
The order of the visual \ac{ECA} and voice combinations was counterbalanced across participants in each block.

\subsubsection*{Statistical analysis}
The procedure for the statistical analysis was identical to that in Experiment 1 except for two changes. First, \ac{IV_Voice} was introduced as one of the \acp{IV}. 
The other two \acp{IV} were the same as in Experiment 1 (\ac{IV_Disp}, and \ac{IV_SP}), with the \acp{DV} also being the same (\textit{Accuracy} and \textit{RT\_mean}). 
Secondly, the priors for \ac{IV_Disp} and \ac{IV_SP} were set based on the respective \ac{PPD} from models in Experiment 1.

\subsection*{Results}
\subsubsection*{\Acl{avVSR}}
Similar to Experiment 1, the LOO criteria was used to determine the final models. For \emph{Accuracy} as the \ac{DV}, the model with all the IVs and an interaction term between \ac{IV_Voice} and \ac{IV_Disp} was selected as the final model since it encompasses the experimental design most comprehensively. The pairwise comparisons for this model are presented in Table~\ref{tab:Exp2_Accuracy}. While the PD for the Monitor condition was greater than 97.5\% (equivalent to a \emph{p}-value of less than .05), suggesting the existence of an effect, neither of the comparisons here were statistically meaningful with the \% of the PPD within ROPE being very high. 

For \emph{RT\_mean} as the DV, the final model also included all the \acp{IV} and an interaction term between \ac{IV_Voice} and \ac{IV_Disp}. However, as in Experiment 1 and for \textit{Accuracy} as the \ac{DV} in this experiment, none of the pairwise comparisons were statistically meaningful, and are not considered further.
\begin{table}[h]
    \centering
    \caption{\bf Summary of pairwise comparisons between the voice incongruence and display device combinations with \textit{Accuracy} as the \ac{DV}.}
\begin{tabular}{|l|l|l|l|l|l|l|}
			\hline
\multicolumn{2}{|l|}{\bf{Comparison}}& \multirow{ 2}{*}{\bf{Median}}    
& \multirow{ 2}{*}{\bf{95\% CI}}           
& \multirow{ 2}{*}{\bf{PD}}      
& \multirow{ 2}{*}{\bf{\% in ROPE}} \\
\cline{1-2}
\bf{Match} & \bf{Device} &&&&\\
\thickhline
        Incong.$-$Cong. & Monitor & 0.02 & [ 0.00, 0.04]  & 97.71\% & 58.72\% \\ 
        Incong.$-$Cong. & \ac{HMD}    & 0.01 & [ 0.00, 0.03]  & 94.80\% & 72.24\% \\ 
        \hline
    \end{tabular}
    \begin{flushleft} CI = Bayesian credible interval, PD = probability of direction, ROPE = Region of practical equivalence, Incong. = Incongruenct, Cong. = Congruent
\end{flushleft}
    \label{tab:Exp2_Accuracy}
\end{table}

\subsubsection*{Questionnaire}
\begin{figure*}[h!]
	\centering
	\includegraphics[width=0.8\textwidth]{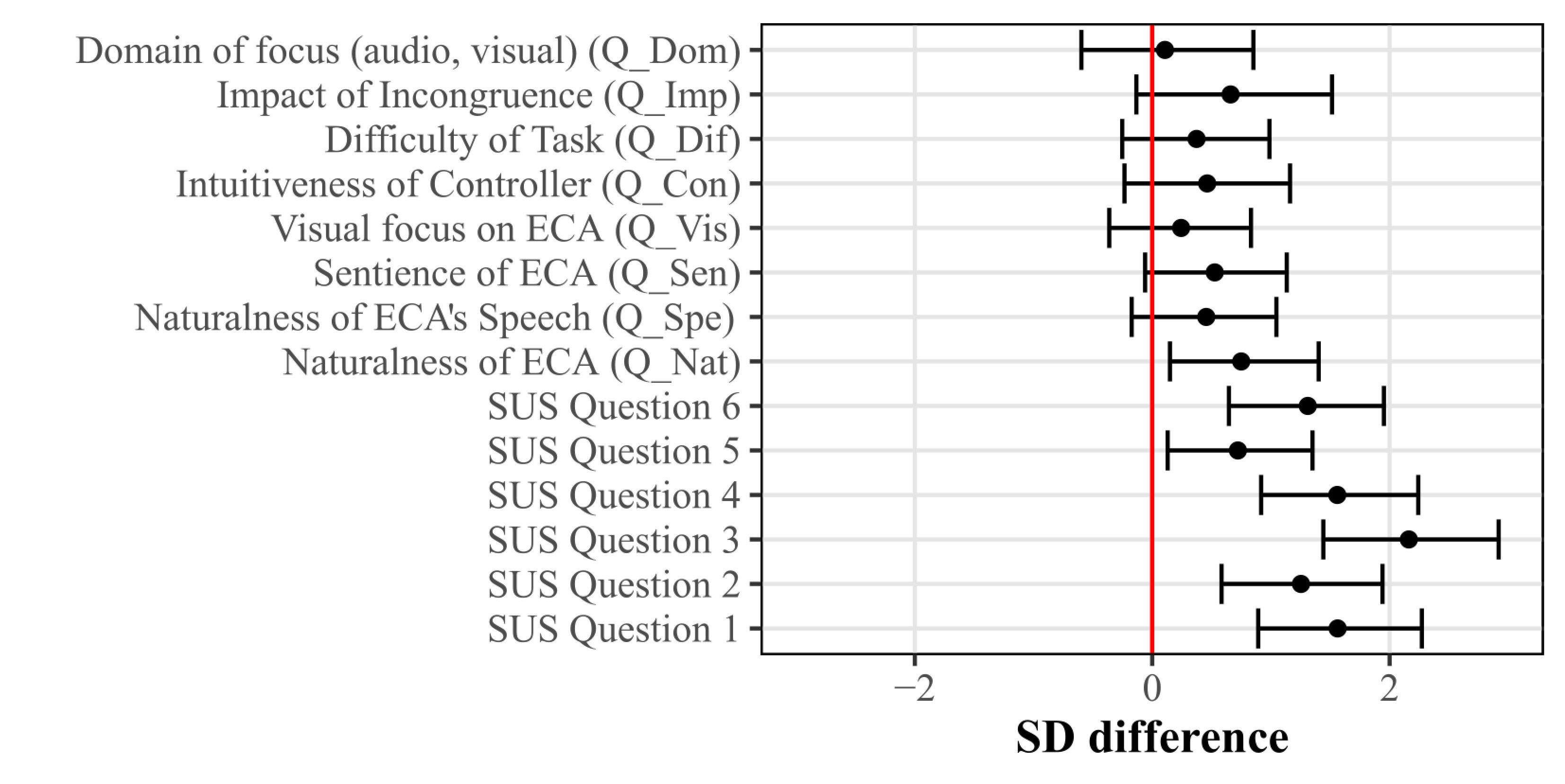}
	\caption{\textbf{Questionnaire results of Experiment 2.} The \emph{SD} difference between the HMD and the Monitor display conditions is shown for each question on the y-axis. Error bars indicate the 95\% credible intervals. All questions were rated on a scale of 1 to 7, excluding {Q\_Inc}, which was asked on a yes/no basis and is thus not displayed here.}
	\label{fig:Exp2_Quest}
\end{figure*}

The results of the questionnaire analysis are displayed in Fig.~\ref{fig:Exp2_Quest}.
As in Experiment 1, the \emph{SD} differences between the ratings on the \ac{SUS} questionnaire for the two display conditions were statistically robust (95\% CI does not include zero), with the participants reporting higher presence in the \ac{HMD} vs. the computer monitor condition. The \acp{ECA}' naturalness (Q\_Nat) was rated higher in the \ac{HMD} condition.

For the remaining questions, the \emph{SD} differences between ratings for the conditions were not significantly robust.  
Only one participant reported not noticing the voice incongruence (Q\_Inc). However, the change in the log odds of the perceivability of the voice incongruence (Q\_Inc) between the monitor and HMD condition (-0.01) was not statistically robust with the 95\% CI ([-4.01, 3.99]) crossing zero. 
This suggests that the perception of incongruence did not vary across the devices.  

\begin{figure*}[h!]
	\centering
	\includegraphics[width=0.8\textwidth]{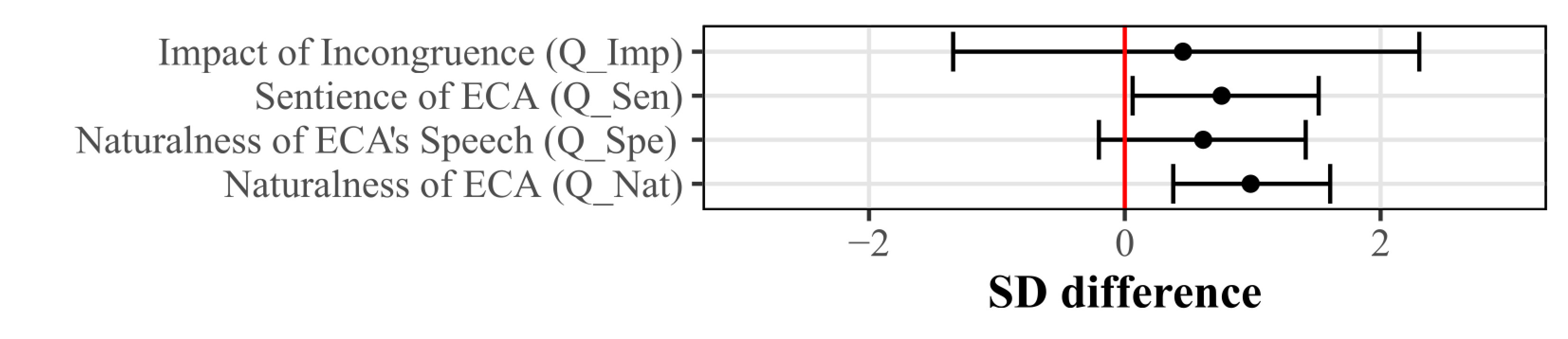}
	\caption{\textbf{\emph{SD} difference of questionnaire data between Experiment 1 and 2.} Results are shown for the questions  Q\_Nat, Q\_Sen, Q\_Spe, and Q\_Imp (on the y-axis). Error bars indicate the 95\% credible intervals. All questions (see Section \textit{Questionnaire}) were rated on a scale of 1 to 7.}
	\label{fig:Exp12_Quest}
\end{figure*}

Finally, the questionnaire results from Experiment 1 were compared with the results from Experiment 2 to see if the changes in the experimental design (changing the type of audiovisual incongruence, adding eye movements to the \ac{ECA} animation) had any effect. 
To that end, the subjective impact of the two types of audiovisual incongruences ({Q\_Imp}) and the three questions related to the \ac{ECA} ({Q\_Nat}, {Q\_Spe}, {Q\_Sen}) were compared between Experiment 1 and Experiment 2 using the same analysis procedure as for the individual questionnaires with the experiment as the \ac{DV}.
The results are displayed in Figure~\ref{fig:Exp12_Quest}.
Indeed, statistically relevant differences in terms of perceived naturalness (Q\_Nat) and sentience (Q\_Sen) were found between Experiments 1 and 2, with higher ratings in Experiment 2.
Differences in the naturalness of speech (Q\_Spe) and the perceived impact of the incongruence (Q\_Imp) were not statistically robust.

\subsection*{Discussion}
In  Experiment 2, task performance in terms of \textit{Accuracy} and \textit{RT\_mean} in the serial recall were examined both for \ac{HMD} and computer monitor presentations.
As in Experiment 1, no influence of the display device on either performance metrics could be found.
This again suggests that the  \ac{avVSR} task can be successfully conducted in \ac{VR}, at least in conditions close to the given setup. 

In this experiment, we aimed to utilize a more obvious and binary audiovisual distractor by introducing a voice incongruence, which reduced the number of participants who failed to notice the incongruence from 13$\%$ in Experiment 1 to 4$\%$ in Experiment 2. 
Even though more people noticed the incongruence, there was no effect on the task performance between the two display conditions, which was the common finding for both the experiments.
In addition, the ratings in Q\_Imp regarding the perceived impact of the incongruence did not differ between the experiments. 

In Experiment 1, a small increase in recall performance could be found for the angle mismatch of $60\si{\degree}$ in the computer monitor presentation. We hypothesize this to be due to the alerting effect of the incongruence. 
In the current experiment, again, the \textit{Accuracy} was slightly increased in trials with incongruent voice and the computer monitor as display device.
However, this finding was not statistically robust.
Even though the type of incongruence was fundamentally different between the experiments, it is reasonable to argue that \textit{Accuracy} is not affected by them in a statistically relevant way.

It is important to note that the current study investigated audiovisual incongruence under a relatively simple condition - specifically, a minimalistic and static laboratory scene featuring only the conversing \acp{ECA} as an additional element. Given this simplicity, our results indicate that it is challenging to detect significant differences in participants' responses. Coherent audiovisual cues are, thus, likely to play a more critical role when focusing on a target source becomes more difficult. To gain further insights into how incongruences affect participants, future research should consider increasing the complexity of the scenes by introducing acoustic distractors (such as background noise) and visual distractors (like additional embodied conversational agents or moving scene objects) that bear increasing resemblance to real-life situations. In a busier scene with multiple distractions, participants may find it increasingly challenging to extract and attend to the voice of an \ac{ECA} if it does not match its visual representation (cp. Experiment 2) or if it perceived from a different direction (cp. Experiment 1).

As in Experiment 1, increased presence could be detected in Experiment 2 for the \ac{HMD} presentation in the questionnaires.
In Experiment 1, the naturalness of speech of the \ac{ECA} was rated higher in the \ac{HMD} condition and the ratings of the general naturalness of the \ac{ECA} was not affected by the display device.
This was the other way around in Experiment 2, where the ratings of the naturalness of speech were not affected by the display device, but the general naturalness of the \ac{ECA} was rated higher in the \ac{HMD} condition, even though the same lip animation algorithm and speech material were used.
This indicates that while the display device does influence the perception of an \ac{ECA}, the specific nature and nuances of this influence remain unclear.
The improved animation of the \ac{ECA} led to better ratings with regard to naturalness and the degree to which the \ac{ECA} is perceived as a sentient being compared to Experiment 1.
Adding even more animations to the \ac{ECA}, like gestures and plausible mimics, could further increase perceived naturalness. 
Since differences in the display device became apparent only in the questionnaire and not in the performance data, our study highlights the need for a multidimensional evaluation framework when assessing \ac{VR} environments. 

\section*{Conclusion}
Two experiments were conducted to investigate the influence of audiovisual angle and voice incongruence, simulated over two display devices (\ac{HMD} and computer monitor), on \ac{avVSR} performance and subjective ratings. With regard to audiovisual incongruences, no clear effect on \textit{Accuracy} or \textit{RT\_mean} could be detected.
Some tendencies of increased performance were observed within the computer monitor presentation for the more extreme angle incongruences.
These tendencies were, however, minor and the same tendencies could not be found for the voice incongruence. 
The only difference between display devices could be found in the \ac{SUS} questionnaire, which consistently revealed higher presence when using an \ac{HMD} compared to a computer monitor in both experiments. 
Ratings of the naturalness and sentience of the \acp{ECA} did not change with the display device.
Hence, our findings suggest that the \ac{avVSR} task can be applied to \ac{VR} settings resembling the current design.

Our findings indicate that the perception of audiovisual incongruences can be applied to more cognitive tasks only to a limited extent, as perceivable incongruences did not affect task performance here.
So far, research on audiovisual incongruence has mostly focussed on the perception level, whereas this paper investigated its relevance for cognitive tasks, which has not been studied previously.
Generally speaking, participants were able to ignore the types of audiovisual incongruences presented in the \ac{avVSR} task.
Further research is needed to explore the relation between perceptibility and task impact in more complex scenes, which could provide further insights into the role of audiovisual congruence in cognitive processing and \ac{VR} design.

\section*{Acknowledgments}
The authors would like to thank Jamilla Balint, Carolin Breuer, Karin Loh, and Julia Seitz for the fruitful discussions, Lukas Vollmer for helping with the analysis, and Alissa Wenzel for assisting with the experiment conduction.

\section*{Data availability}
\label{sec:data}
The following plugins by the Virtual Reality \& Immersive Visualization Group at RWTH Aachen University were used to implement this project:
\begin{itemize}
	\item RWTH VR Toolkit (for basic \ac{VR} interaction) \url{https://git-ce.rwth-aachen.de/vr-vis/VR-Group/unreal-development/plugins/rwth-vr-toolkit}, see also Gilbert et al.~\cite{gilbert_2024_10817754} 
	\item Character Plugin (Animation of MetaHuman Lip Movement and Eye Gazing) \url{https://git-ce.rwth-aachen.de/vr-vis/VR-Group/unreal-development/plugins/character-plugin}
	\item Study Framework Plugin (Control of listening experiment, logging of data) \url{https://git-ce.rwth-aachen.de/vr-vis/VR-Group/unreal-development/plugins/unreal-study-framework}, see also Ehret et al.~\cite{ehret.StudyFrameworkComfortablySetting.2024}
	\item Virtual Acoustics Plugin (Communication with \ac{VA} server) \url{https://git-ce.rwth-aachen.de/vr-vis/VR-Group/unreal-development/plugins/unreal-va-plugin}, see also Schäfer et al.~\cite{institute_for_hearing_technology_and_aco_2024_13744474}
\end{itemize}

\section*{Author contributions statement}
C.E. and J.F. conceived the experiments, C.E. and J.E. implemented the experiments, C.E. conducted the experiments, M.Y., C.M., and C.E. analyzed the results. C.E. and M.Y. interpreted the results supported by J.E., A.B., C.M., and S.S.
C.E.  took the lead in writing the manuscript and was supported by M.Y. and J.F. All authors provided critical feedback and helped finalize the research, analyses, and manuscript. 
J.F. supervised the project and provided all necessary resources.

\end{document}